\documentclass[twocolumn,showpacs,preprintnumbers,amsmath,amssymb]{revtex4}

\usepackage{graphicx}
\usepackage{dcolumn}
\usepackage{bm}

\begin{document}

\title{Comment on: ``Self-Diffusion in 2D Dusty-Plasma Liquids: Numerical-Simulation Results''}

\author{T. Ott}\author{M. Bonitz}%
\affiliation{%
    Christian-Albrechts-Universit\"at zu Kiel, Institut f\"ur Theoretische Physik und Astrophysik, Leibnizstra\ss{}e 15, 24098 Kiel, Germany
}%
\author{P. Hartmann}
\affiliation{
Research Institute for Solid State Physics and Optics, Hungarian Academy of Sciences, P. O. Box 49, H-1525 Budapest, Hungary
}%

\date{\today}
 \maketitle
\
In a recent Letter~\cite{hou:085002}, Hou \textit{et al.}~(HPS) presented numerical results for the diffusion 
process in two-dimensional dusty plasma liquids with Yukawa pair interaction [2DYL], $V(r)=Q^2 \exp(-r/\lambda)/r$, by solving a Langevin equation.
The mean-squared displacement
\begin{equation}
 u_r(t)=\langle \vert \vec r(t) - \vec r(t_0) \vert^2 \rangle \propto t^{1+\alpha}\label{eq:diag_msd}\,
\end{equation}
is used to distinguish normal diffusion ($\alpha=0$) from subdiffusion ($\alpha<0$) and superdiffusion ($\alpha>0$). 
HPS observed superdiffusion and reported a complicated non-monotonic dependence of $\alpha$ on the potential stiffness $\kappa=a/\lambda$, where $a$ is the mean interparticle distance. 
Here we point out that the {\em behavior} $\alpha(\kappa)$ {\em is, in fact, regular and systematic}, whereas the observations of Ref. \cite{hou:085002} resulted from a comparison of different system states.

As noted in \cite{hou:085002}, $\alpha$ depends on $\kappa$ and the coupling parameter 
$\Gamma=(Q^2/4\pi\varepsilon_0)\times(1/ak_BT)$ and finding the dependence $\alpha(\kappa)$ requires to compare states with the same physical coupling. This can be done by fixing, for all $\kappa$, the value $\Gamma^{\rm rel}=\Gamma/\Gamma_c$, where $\Gamma_c(\kappa)$ is the crystallization point which is well known for 
$\kappa\leq 3$~\cite{Hartmann2005}. For larger $\kappa$, we obtain 
$\Gamma_c(\kappa=3.5)=2340$ and  $\Gamma_c(4)=4500$. 

\begin{figure}[h]
 \centering
 \includegraphics[scale=0.7]{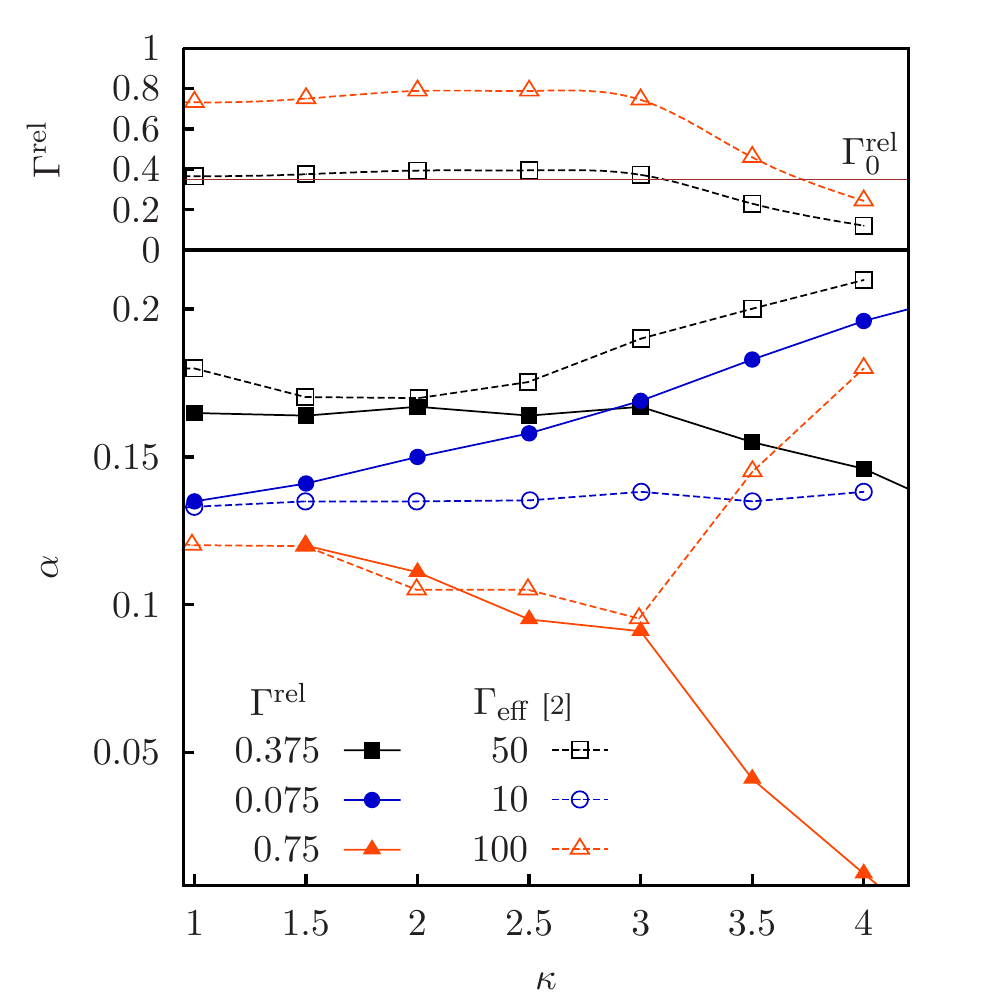}
 \caption{Bottom: Exponent $\alpha$ vs. $\kappa$ for three fixed values of $\Gamma^{\rm rel}$ (full lines and symbols) and $\Gamma_{\rm eff}$ (dashed lines, open symbols, data from Ref.~\cite{hou:085002}). Top: $\Gamma^{\rm rel}(\kappa)$ corresponding to the values $\Gamma_{\rm eff}$ used in \cite{hou:085002}. 
}
 \label{im1}
\end{figure}

We have performed detailed investigations of the dependence of $\alpha$ on $\Gamma$ and $\kappa$
\cite{ott_dipl} and observed two different regimes: i) for $\Gamma^{\rm rel} \lesssim \Gamma^{\rm rel}_0 = 0.35$, 
$\alpha$ is monotonically decreasing with $\kappa$, at constant $\Gamma^{\rm rel}$. 
ii) for $\Gamma^{\rm rel} \gtrsim \Gamma^{\rm rel}_0$, $\alpha$ increases monotonically with $\kappa$, at constant $\Gamma^{\rm rel}$. 
Around $\Gamma^{\rm rel} = \Gamma^{\rm rel}_0$, $\alpha$ is almost independent of $\kappa$. 
Fig.~\ref{im1} clearly confirms the monotonic $\kappa$-dependence of $\alpha$ for three fixed values of $\Gamma^{\rm rel}$ corresponding to the parameters shown in Fig. 5 of \cite{hou:085002}.

HPS used a different coupling parameter, $\Gamma_{\rm eff}$, which yields an almost constant $\Gamma^{\rm rel}$, for $\kappa \le 3$. However, for $\kappa > 3$ 
it corresponds to strongly varying $\Gamma^{\rm rel}$ and thus to different physical situations, \cite{fit}, cf. top part of Fig.~\ref{im1}. For example, their value, $\Gamma_{\rm eff}=100$, corresponds to $\Gamma^{\rm rel}=0.76>\Gamma^{\rm rel}_0$, for $\kappa = 3$, but to $\Gamma^{\rm rel}=0.24 < \Gamma^{\rm rel}_0$, for $\kappa=4$. 
This explains the non-monotonicity of $\alpha(\kappa)$ reported by HPS \cite{time}.

Thus, we report a {\em systematic effect of screening} on superdiffusion in 2DYL based on numerical simulations. An increase of $\kappa$ 
supports superdiffusion for $\Gamma^{\rm rel} \lesssim 0.6\cdot \Gamma^{\rm rel}_0$ and results in an increasing diffusion exponent 
in this range of the coupling. For higher couplings, $\Gamma^{\rm rel} \gtrsim 0.6\cdot\Gamma^{\rm rel}_0$, a stronger screening 
has the inverse effect and reduces the strength of anomalous diffusion. In conclusion, we have presented numerical evidence for 
the existence of a monotonic dependence of anomalous diffusion on screening. An explanation of this behaviour is beyond the present Comment and will be given elsewhere.

\bibliographystyle{apsrev}

\begin{thebibliography}{5}
\bibitem{time} 
The different absolute values of our $\alpha$, for $\kappa \le 3$, compared to HPS are most likely due to a different prescription for extracting $\alpha$ from $u_r(t)$. We have always used a constant time interval $\omega_pt\in [100,320]$ to read off the slope of $u_r(t)$. A friction coefficient $\nu/\omega_p=0.001$ was used, as in \cite{hou:085002}.
\end{thebibliography}

\end{document}